\documentclass[aps,prb,twocolumn,superscriptaddress,showpacs,showkeys,floatfix]{revtex4}

\usepackage{graphicx,color}

\newcommand{\jwj}[1]{\textcolor{red}{#1}}

\graphicspath{{figs/}}
\begin{document}
\title{MoS$_{2}$ Nanoresonators: Intrinsically Better Than Graphene?}
\author{Jin-Wu Jiang}
    \altaffiliation{Corresponding author: jwjiang5918@hotmail.com}
    \affiliation{Shanghai Institute of Applied Mathematics and Mechanics, Shanghai Key Laboratory of Mechanics in Energy Engineering, Shanghai University, Shanghai 200072, People's Republic of China}
    \affiliation{Institute of Structural Mechanics, Bauhaus-University Weimar, Marienstr. 15, D-99423 Weimar, Germany}
\author{Harold S. Park}
    \altaffiliation{Corresponding author: parkhs@bu.edu}
    \affiliation{Department of Mechanical Engineering, Boston University, Boston, Massachusetts 02215, USA}
\author{Timon Rabczuk}
    \altaffiliation{Corresponding author: timon.rabczuk@uni-weimar.de}
    \affiliation{Institute of Structural Mechanics, Bauhaus-University Weimar, Marienstr. 15, D-99423 Weimar, Germany}
    \affiliation{School of Civil, Environmental and Architectural Engineering, Korea University, Seoul, South Korea }

\date{\today}
\begin{abstract}
We perform classical molecular dynamics simulations to examine the intrinsic energy dissipation in single-layer MoS$_{2}$ nanoresonators, where a point of emphasis is to compare its dissipation characteristics with those of single-layer graphene.  Our key finding is that MoS$_{2}$ nanoresonators exhibit significantly lower energy dissipation, and thus higher quality (Q)-factors by at least a factor of four below room temperature, than graphene.  Furthermore, this high Q-factor endows MoS$_{2}$ nanoresonators with a higher figure of merit, defined as frequency times Q-factor, despite a resonant frequency that is 50\% smaller than graphene for the same size.  By utilizing arguments from phonon-phonon scattering theory, we show that this reduced energy dissipation is enabled by the large energy gap in the phonon dispersion of MoS$_{2}$, which separates the acoustic phonon branches from the optical phonon branches, leading to a preserving mechanism for the resonant oscillation of MoS$_{2}$ nanoresonators.  We further investigate the effects of tensile mechanical strain and nonlinear actuation on the Q-factors, where the tensile strain is found to counteract the reductions in Q-factor that occur with higher actuation amplitudes. Overall, our simulations illustrate the potential utility of MoS$_{2}$ for high frequency sensing and actuation applications.

\end{abstract}

\maketitle
\pagebreak

\section{Introduction}
Graphene nanoresonators are a promising candidate for ultrasensitive mass sensing and detection due to its desirable combination of high stiffness and large surface area.\cite{BunchJS2007sci,LeeC2008sci,JiangJW2009young,JiangJW2010young,EkinciKL,ArletJL,EomK} For such sensing applications, it is important that the nanoresonator exhibit low energy dissipation, or a high quality (Q)-factor, since the sensitivity of the nanoresonator is inversely proportional to its Q-factor\cite{EkinciKL}.  Various energy dissipation mechanisms have been explored in graphene nanoresonators, such as external attachment energy loss,\cite{SeoanezC,KimSY2009apl} intrinsic nonlinear scattering mechanisms,\cite{AtalayaJ} the effective strain mechanism,\cite{JiangJW2012nanotechnology} edge effects,\cite{KimSY2009nl,JiangJW2012jap}, grain boundary-mediated scattering losses\cite{QiZ2012nns}, and the adsorbate migration effect.\cite{JiangJW2013diffusion}

Thanks to the recent experimental improvements in producing large-area, highly-pure samples of single and few-layer graphene, there has recently appeared several interesting experimental studies on graphene nanoresonators.  For example, at low temperature, extremely weak energy dissipation was observed in very pure graphene nanoresonators.\cite{EichlerA,ZandeAMVD,ChenC2009nn} However, these experiments also show that the energy dissipation increases substantially with increasing temperature for graphene nanoresonators, where theoretically it was found that the Q-factors decrease according to a $1/T$ scaling.\cite{KimSY2009nl}  Furthermore, many sensing applications are expected to occur at temperatures approaching room temperature, and thus it is important and of practical significance to examine if other two-dimensional materials exhibit less intrinsic energy dissipation, which would be highly beneficial for applications that depend on two-dimensional nanoresonators.

Another two-dimensional material that has recently gained significant interest is Molybdenum Disulphide (MoS$_{2}$).  The primary reason for the excitement surrounding MoS$_{2}$ is due to its superior electronic properties as compared to graphene, starting with the fact that in bulk form it exhibits a band gap of around 1.2~{eV},\cite{KamKK} which can be further increased by changing its thickness,\cite{MakKF} or through application of mechanical strain.\cite{FengJ2012npho,LuP2012pccp} This finite band gap is a key reason for the excitement surrounding MoS$_{2}$ as compared to graphene due to the well-known fact that graphene is gapless.\cite{NovoselovKS2005nat}  Because of its direct bandgap and also its well-known properties as a lubricant, MoS$_{2}$ has attracted considerable attention in recent years.\cite{WangQH2012nn,ChhowallaM} For example, Radisavljevic et al.\cite{RadisavljevicB2011nn} demonstrated the potential of single-layer MoS$_{2}$ as a transistor. The strain and the electronic noise effects were found to be important for single-layer MoS$_{2}$ transistors.\cite{ConleyHJ,SangwanVK,Ghorbani-AslM,CheiwchanchamnangijT} Besides the electronic properties, there has also been increasing interest in the thermal and mechanical properties of mono and few-layer MoS$_{2}$.\cite{HuangW,VarshneyV,BertolazziS,CooperRC2013prb1,CooperRC2013prb2,JiangJW2013mos2,JiangJW2013sw,JiangJW2013bend}

Very recently, two experimental groups have demonstrated the nanomechanical resonant behavior for single-layer MoS$_{2}$\cite{Castellanos-GomezA2013adm} or few-layer MoS$_{2}$\cite{LeeJ2013acsnano}. Interestingly, Lee et. al found that MoS$_{2}$ exhibits a higher figure of merit, i.e. frequency-Q-factor product $f_{0}\times Q\approx10^{10}$ Hz, than graphene.\cite{LeeJ2013acsnano}  While this experiment intriguingly suggests that MoS$_{2}$ may exhibit lower intrinsic energy dissipation than graphene, a systematic theoretical investigation and explanation for this fact is currently lacking.  Therefore, the aim of the present work is to examine the intrinsic energy dissipation in MoS$_{2}$ nanoresonators, with comparison to graphene.

In doing so, we report for the first time that MoS$_{2}$ nanoresonators exhibit significantly less intrinsic energy dissipation, and also a higher figure of merit, than graphene. Furthermore, we find that the origin for this reduced energy dissipation is the large energy gap in the phonon dispersion of MoS$_{2}$, which helps prevent the resonant oscillation from being deleteriously affected by other phonon modes. We also find that the energy dissipation in both MoS$_{2}$ and graphene nanoresonators is considerably enhanced when larger actuation amplitudes are prescribed due to the emergence of ripples.  However, we also show that these ripples can be removed and the enhanced energy dissipation can be mitigated through the application of tensile mechanical strain.

\section{Structure and simulation details}
The single-layer graphene and single-layer MoS$_{2}$ samples in our simulation are 200~{\AA} in the longitudinal direction and 20~{\AA} in the lateral direction. The interaction between carbon atoms in graphene is described by the Brenner (REBO-II) potential~\cite{brennerJPCM2002}. The interaction within MoS$_{2}$ is described by the recently developed Stillinger-Weber potential.\cite{JiangJW2013sw} The standard Newton equations of motion are integrated in time using the velocity Verlet algorithm with a time step of 1 fs. Both ends in the longitudinal direction are fixed while periodic boundary conditions are applied in the lateral direction.

Our simulations are performed as follows.  First, the Nos\'e-Hoover\cite{Nose,Hoover} thermostat is applied to thermalize the system to a constant temperature within the NPT (i.e. the number of particles N, the pressure P and the temperature T of the system are constant) ensemble, which is run for 100~{ps}.  Free mechanical oscillations of the nanoresonators are then actuated by adding a sine-shaped velocity distribution to the system in the $z$ direction~\cite{JiangJW2012jap}, where the $z$ direction is perpendicular to the graphene or MoS$_{2}$ plane. The imposed velocity for atom $i$ is $\vec{v}_{\rm i}=\beta \sin (\pi x_{\rm i}/L)\vec{e}_{\rm z}$. \jwj{The actuation parameter $
\beta$ determines the resonant oscillation amplitude, $A=\beta/\omega$, where $\omega=0.25$ or 0.5~{ps$^{-1}$} is the angular frequency for the graphene and MoS$_{2}$ nanoresonators in present work. For $\beta=1.0$~{\AA ps$^{-1}$}, the resonant oscillation amplitude is 2~{\AA} for graphene and 4.0~{\AA} for MoS$_{2}$, which is only 1\% or 2\% of the length of the resonator. The corresponding effective strain is 0.037\% for MoS$_{2}$.\cite{JiangJW2012nanotechnology}}

\begin{figure}[htpb]
  \begin{center}
    \scalebox{1}[1]{\includegraphics[width=8cm]{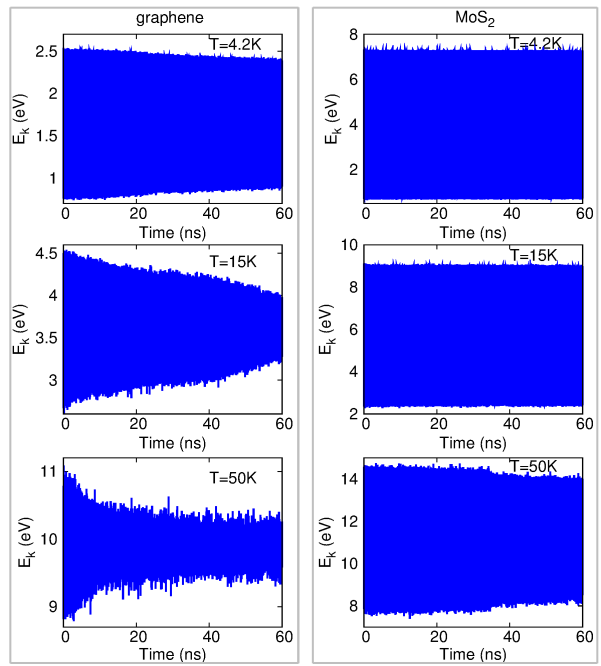}}
  \end{center}
  \caption{(Color online) Kinetic energy time history in graphene (left) and MoS$_{2}$ (right) nanoresonators at different temperatures. The actuation parameter $\beta=2$ for all calculations here. Left: the energy dissipation in graphene nanoresonators increases quickly with increasing temperature. Right: the energy dissipation in MoS$_{2}$ nanoresonators increases slowly with increasing temperature, and thus the MoS$_{2}$ nanoresonator exhibits a lower intrinsic energy dissipation than graphene nanoresonator at the same temperature.}
  \label{fig_energy_temperature}
\end{figure}

\begin{figure}[htpb]
  \begin{center}
    \scalebox{1}[1]{\includegraphics[width=8cm]{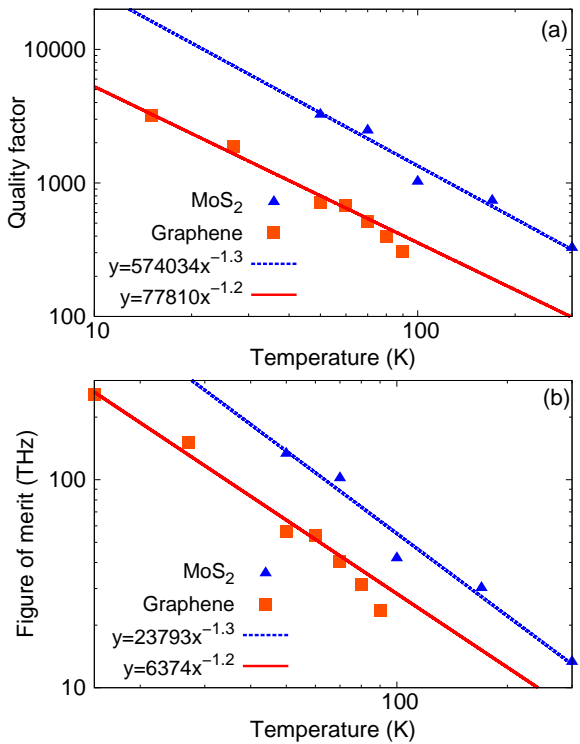}}
  \end{center}
  \caption{(Color online) Temperature dependence of the (a) Q factor and (b) figure of merit (i.e $f_{0}\times Q$)  in graphene and MoS$_{2}$ nanoresonators. The actuation parameter $\beta=2$ for all calculations here.}
  \label{fig_qfactor}
\end{figure}

\begin{figure}[htpb]
  \begin{center}
    \scalebox{0.95}[0.95]{\includegraphics[width=8cm]{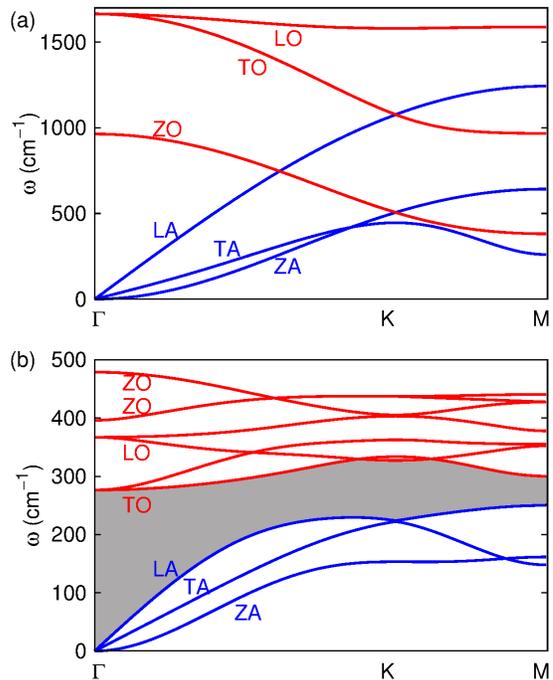}}
  \end{center}
  \caption{(Color online) Phonon dispersion of graphene and MoS$_{2}$ along the high symmetry $\Gamma$KM lines in the Brillouin zone. (a) Phonon dispersion for graphene calculated from the Brenner potential. Note the crossing of the acoustic and optical branches. (b) Phonon dispersion for MoS$_{2}$ calculated from the Stillinger-Weber potential.  Note the clear energy gap (gray area) between the acoustic and optical branches, i.e there is no cross-over between the acoustic and optical branches.}
  \label{fig_dispersion}
\end{figure}

\begin{figure}[htpb]
  \begin{center}
    \scalebox{0.95}[0.95]{\includegraphics[width=8cm]{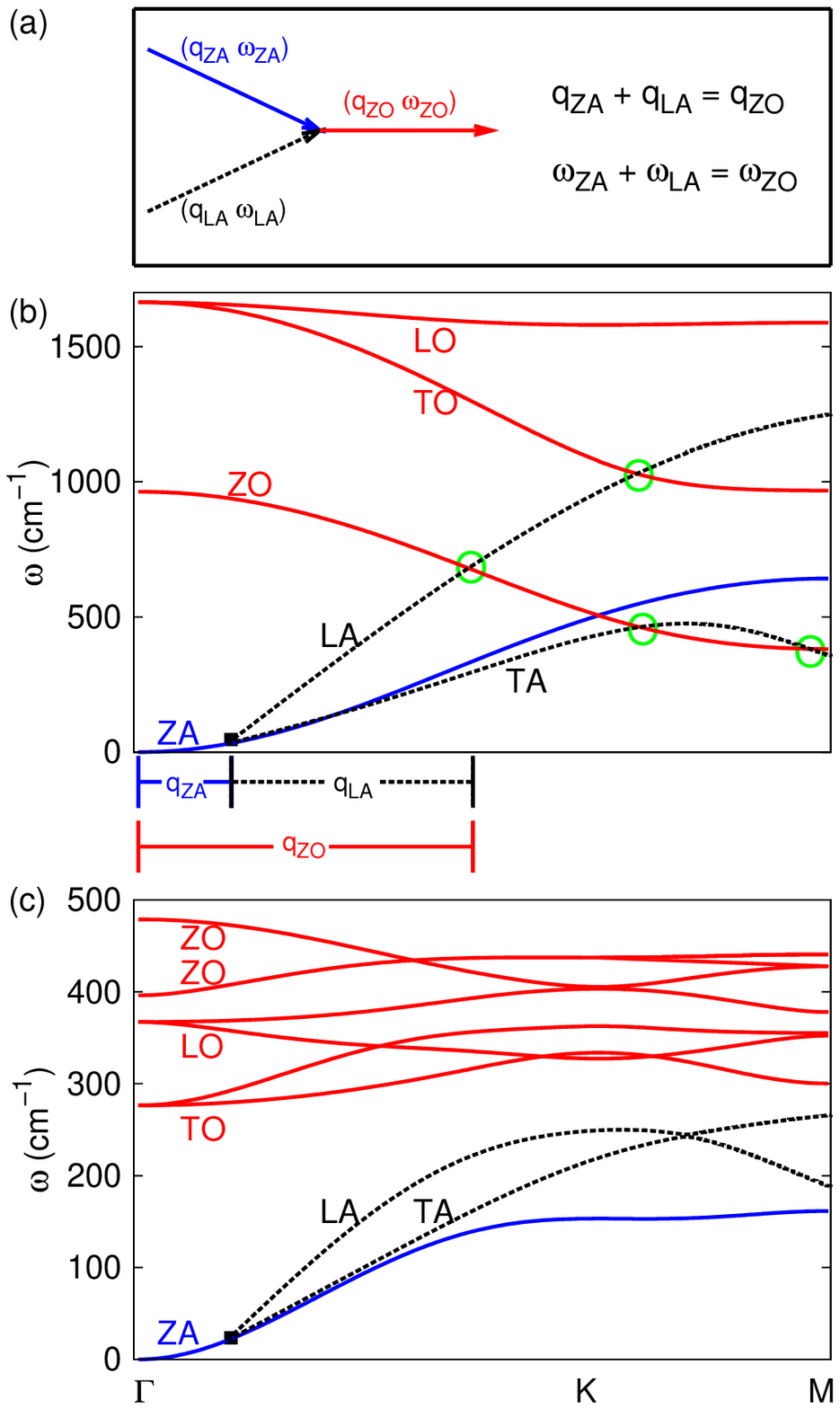}}
  \end{center}
  \caption{(Color online) Illustration of the phonon-phonon scattering mechanism in the graphene and MoS$_{2}$ nanoresonators. (a) A typical scattering process. The ZA mode is scattered by the other acoustic LA mode. As a result of this phonon-phonon scattering, one optical ZO mode is created.  The energy and momentum constraints are: $q_{\rm ZA}+q_{\rm LA}=q_{\rm ZO}$ and $\omega_{\rm ZA} + \omega_{\rm LA}=\omega_{\rm ZO}$. (b) The phonon-phonon scattering of a low-frequency ZA mode $(q_{\rm ZA}, \omega_{\rm ZA})$ in graphene, where both energy and momentum constraints are satisfied. The origins of the TA and LA branches (black dashed lines) are shifted to the position of this ZA mode.  There are four cross-over points (green circles) between TA/LA and the optical branches (red solid lines), which correspond to four permitted phonon-phonon scattering processes. (c) There is no cross-over between the shifted TA/LA branches and the optical branches in MoS$_{2}$, because of the energy gap between the acoustic and optical branches.}
  \label{fig_scattering}
\end{figure}

After the actuation of the mechanical oscillation, the system is allowed to oscillate freely within the NVE (i.e. the particles number N, the volume V and the energy E of the system are constant) ensemble. The data from this NVE ensemble is used to analyze the mechanical oscillation of the nanoresonators. All molecular dynamics simulations were performed using the publicly available simulation code LAMMPS~\cite{PlimptonSJ,Lammps}, while the OVITO package was used for visualization~\cite{ovito}.

\section{Results and discussions}
\subsection{Lower Intrinsic Energy Dissipation in MoS$_{2}$ than Graphene Nanoresonators}

We first compare the temperature dependence for the intrinsic energy dissipation in the single-layer graphene and MoS$_{2}$ nanoresonators. Fig.~\ref{fig_energy_temperature} shows the kinetic energy time history in both graphene (left) and MoS$_{2}$ (right) nanoresonators when the actuation energy parameter $\beta=2.0$. The oscillation amplitude in the kinetic energy decays gradually, which reflects the dissipation of the resonant oscillation energy within the nanoresonator. A common feature in the graphene\cite{KimSY2009nl} and MoS$_{2}$ nanoresonators is that the energy dissipation becomes larger with increasing temperature. 

Fig.~\ref{fig_qfactor}~(a) shows the Q-factor extracted from the kinetic energy time history shown in Fig.~\ref{fig_energy_temperature} for both graphene and MoS$_{2}$. The decay of the oscillation amplitude for the kinetic energy is used to extract the Q-factor by fitting the kinetic energy from the NVE ensemble to a function $E_{k}(t)=a+b(1-2\pi/Q)^{t} \cos(\omega t)$, where $\omega$ is the frequency, $a$ and $b$ are two fitting parameters and $Q$ is the resulting quality factor.\cite{JiangJW2012jap} The Q-factor in MoS$_{2}$ is clearly higher than that of graphene, and is greater by at least a factor of four for all temperatures below room temperature. In particular, the Q-factor at room temperature for MoS$_{2}$ is 327 from our simulations, which is much higher than the Q-factor of 83 in graphene as extrapolated from the fitting formula. Fig.~\ref{fig_qfactor}~(a) also shows that the Q-factors for single-layer MoS$_{2}$ decay with temperature according to a $Q\sim1/T^{-1.3}$ relationship, similar to the $Q\sim1/T^{-1.2}$ we find for single-layer graphene, where the $T^{-1.2}$ relationship we report is slightly different than the $T^{-1}$ relationship found previously by Kim and Park\cite{KimSY2009nl} due to differences in how the Q-factor was calculated.
 
Fig.~\ref{fig_qfactor}~(b) compares the figure of merit (i.e. $f_{0}\times Q$), in graphene and MoS$_{2}$ nanoresonators. The figure of merit in MoS$_{2}$ is also higher than that  for graphene although the frequency for MoS$_{2}$ (40~GHz) is only half of the frequency for graphene (80~GHz), which again is due to the substantially higher Q-factors for MoS$_{2}$ nanoresonators.

To understand the energy dissipation in graphene and MoS$_{2}$ nanoresonators, we shall analyze the relationship between the mechanical resonant oscillation in the nanoresonator and the phonon modes in the lattice dynamics theory. The resonant oscillation of these two-dimensional structures is actually the mechanical vibration of their out of plane (z)-direction acoustic (ZA) modes, so the only energy dissipation mechanism here is due to phonon-phonon scattering.\cite{JiangJW2010epl} This ZA mode is scattered by other phonon modes, which have higher density of states at higher temperature. As a result, the scattering of the ZA mode becomes stronger with increasing temperature. This is the origin for the increase in energy dissipation with increasing temperature observed in Fig.~\ref{fig_energy_temperature}. \jwj{It should be noted that boundary scattering does not play a role here, because there is no temperature gradient in the simulation of the resonant oscillation. The system has been thermalized to a constant temperature within the NPT ensemble prior to the actuation of the mechanical oscillation. As a result, the flexural phonons are not transported to the boundary of the system. Instead, the flexural phonon acts as a stationary mode in the system.}

From Fig.~\ref{fig_energy_temperature}, it is obvious that the MoS$_{2}$ nanoresonator exhibits much smaller energy dissipation than the graphene nanoresonator at the same temperature, where the difference becomes more distinct with increasing temperature.

To reveal the underlying mechanism for this difference, we first identify and discuss some fundamental phonon modes in graphene and MoS$_{2}$, because phonon-phonon scattering is the only energy dissipation mechanism that is operant here. Specifically, there are three acoustic branches, i.e the ZA branch, the transverse acoustic (TA) branch, and the longitudinal acoustic (LA) branch. There are also three optical branches, i.e the z-direction optical (ZO) branch, the transverse optical (TO) branch, and the longitudinal optical (LO) branch.

Fig.~\ref{fig_dispersion} shows the phonon dispersion of single-layer graphene and MoS$_{2}$ along the high symmetry $\Gamma$KM lines in the Brillouin zone. Fig.~\ref{fig_dispersion}~(a) is the phonon dispersion for graphene calculated from the Brenner potential.\cite{brennerJPCM2002} The three acoustic branches are plotted by blue solid lines. The three optical branches are plotted by red solid lines.  The key feature is the fact that the acoustic and optical branches exhibit a cross-over, where this cross-over is a general feature in the phonon dispersion curves for single-layer graphene obtained from different methods, e.g. the force constant model,\cite{SaitoR} first-principles calculations,\cite{MaultzschJ2004prl} or experiments.\cite{AizawaT,MaultzschJ2004prl,MohrM2007prb} 

Fig.~\ref{fig_dispersion}~(b) is the phonon dispersion for single-layer MoS$_{2}$ calculated from the Stillinger-Weber potential.\cite{JiangJW2013sw} In contrast to graphene, there is an energy gap (gray area) between the acoustic and optical branches, i.e there is no cross-over between the acoustic and optical branches. This is again a general feature in the phonon dispersion curves of MoS$_{2}$ obtained from different methods, e.g. force constant model,\cite{JimenezSS,DamnjanovicM2008mmp} first-principles calculations,\cite{SanchezAM} or experiments.\cite{WakabayashiN}  The key effect of this energy gap is to separate the acoustic phonon branches from the optical phonon branches in single-layer MoS$_{2}$.

\jwj{The strength of the phonon-phonon scattering is simultaneously determined by two aspects. First, it is proportional to the square of the nonlinear elastic constant. The nonlinear elastic constant in MoS$_{2}$ is about -1.8~{TPa} from the Stillinger-Weber potential used in our simulation, which is close to the value of -2.0~{TPa} in graphene.\cite{LeeC2008sci,JiangJW2010young} Secondly, the symmetry selection rule plays a key role in determining the strength of the phonon-phonon scattering.} In the phonon-phonon scattering mechanism, the symmetric selection rule requires phonon modes from different branches to be involved.\cite{BornM,Holland1963pr} A typical scattering process is shown in Fig.~\ref{fig_scattering}~(a), where the ZA mode is scattered by the other acoustic modes LA (or TA). As a result of this phonon-phonon scattering, another optical mode ZO (or LO, or TO) is created. The energy and momentum conservation laws add two strict constraints on the phonon-phonon scattering process, i.e. $q_{\rm ZA}+q_{\rm LA}=q_{\rm ZO}$ and $\omega_{\rm ZA} + \omega_{\rm LA}=\omega_{\rm ZO}$. We note that this corresponds to the Normal phonon-phonon scattering process. The Umklapp scattering is another phonon-phonon scattering process, where the momentum conservation is relaxed by allowing the appearance of a reciprocal lattice vector.\cite{BornM} Our discussions here are also applicable for the Umklapp process. 

\begin{figure}[htpb]
  \begin{center}
    \scalebox{1}[1]{\includegraphics[width=8cm]{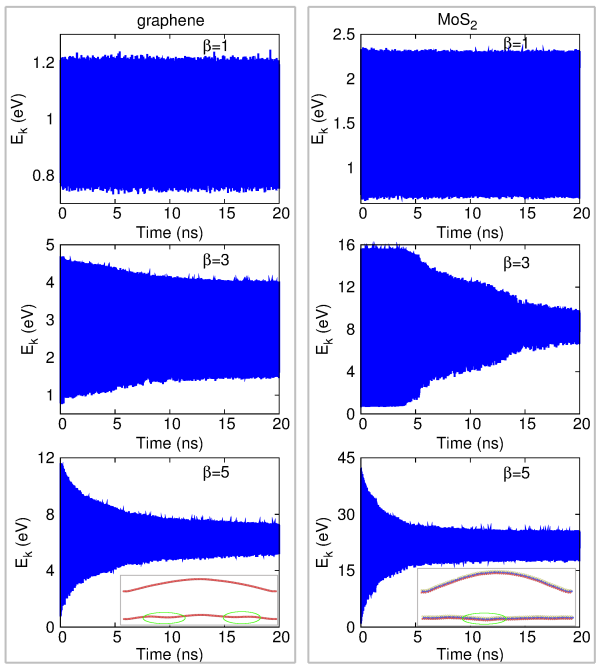}}
  \end{center}
  \caption{(Color online) Kinetic energy time history for graphene (left) and MoS$_{2}$ (right) nanoresonators with different actuation parameters $\beta$. Temperature T=4.2~K for this set of calculations. In both graphene and MoS$_{2}$ nanoresonators, the energy dissipation increases with increasing actuation parameter $\beta$. Insets show two special configurations from early stage of molecular dynamics simulation, which correspond to minimum or maximum kinetic energy. Ripples are indicated by green ellipses.}
  \label{fig_energy_alpha}
\end{figure}

\begin{figure}[htpb]
  \begin{center}
    \scalebox{1}[1]{\includegraphics[width=8cm]{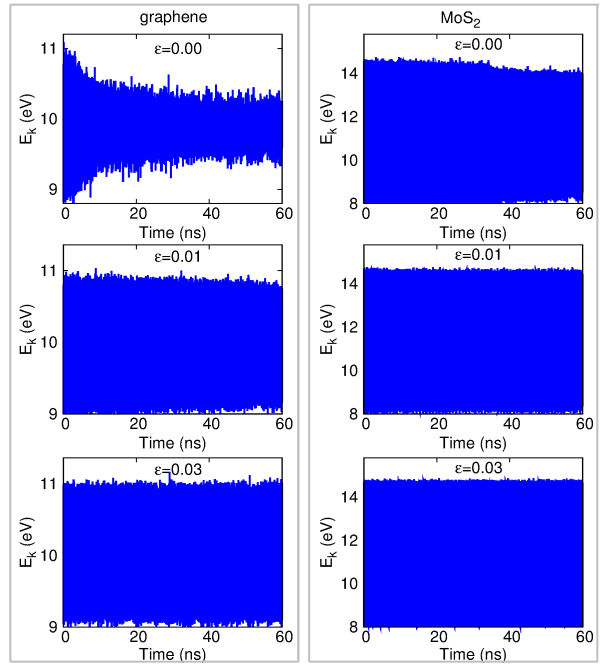}}
  \end{center}
  \caption{(Color online) Kinetic energy time history for graphene (left) and MoS$_{2}$ (right) nanoresonators with different mechanical tension $\epsilon$. Temperature T=50~K and actuation parameter $\beta=2.0$ for this set of simulations. The mechanical strain strongly alleviates the energy dissipation in graphene nanoresonators, while having a less pronounced effect on MoS$_{2}$, though the intrinsic dissipation in MoS$_{2}$ is much smaller than graphene.}
  \label{fig_energy_strain}
\end{figure}

\begin{figure*}[htpb]
  \begin{center}
    \scalebox{0.85}[0.85]{\includegraphics[width=\textwidth]{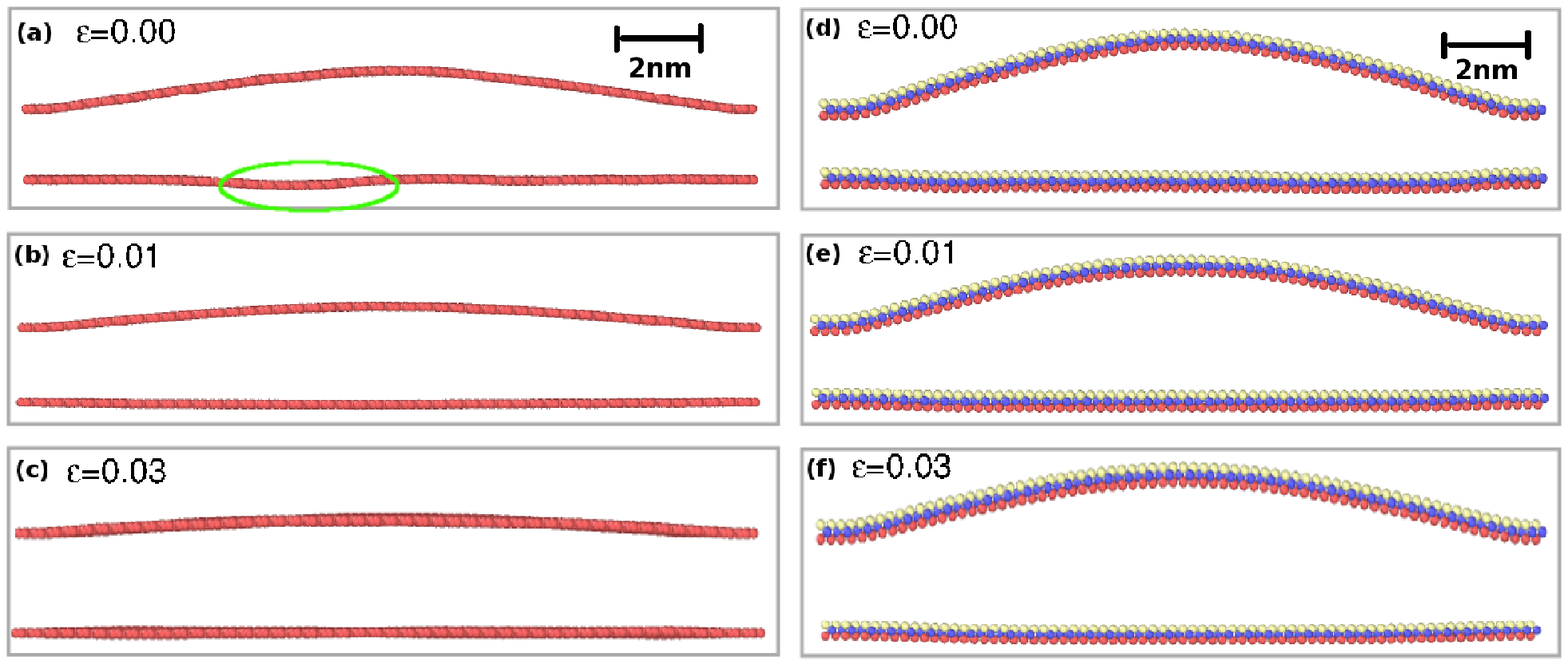}}
  \end{center}
  \caption{(Color online) Configurations for graphene and MoS$_{2}$ nanoresonators from an early stage of the molecular dynamics simulation at T=50.0~K and $\beta=2.0$. (a)-(c) are configurations of graphene nanoresonators with increasing mechanical tension. Two special configurations, which correspond to minimum or maximum kinetic energy, are shown. In (a), some ripples (indicated by green ellipses) can be found in the (horizontal) configuration with maximum kinetic energy, as a result of the thermal vibration of the graphene. These ripples in the graphene nanoresonator can be removed by mechanical tension as shown in (b) and (c). (d)-(f) are configurations for MoS$_{2}$ nanoresonators with increasing mechanical tension. Ripples in the MoS$_{2}$ nanoresonator are not present due to large bending modulus of MoS$_{2}$.}
  \label{fig_cfg_strain}
\end{figure*}

Fig.~\ref{fig_scattering}~(b) shows the phonon-phonon scattering of a low-frequency ZA mode in graphene, where both energy and momentum constraints are satisfied. We have chosen a particular ZA mode with $(q_{\rm ZA}, \omega_{\rm ZA})$ from the ZA branch (blue solid line). The origins of the TA and LA branches (black dashed lines) are shifted to the position of this ZA mode. In this way, the crossing point between TA/LA and the optical branches will disclose all permitted phonon-phonon scattering processes (i.e with conserved energy and momentum). There are four cross-over points (green circles) between TA/LA and the optical branches (red solid lines). These four crossing points correspond to four permitted phonon-phonon scattering processes in single-layer graphene. In the horizontal axis, we have depicted the corresponding wave vectors of the three phonon modes for a crossing between the LA and ZO branches, i.e. $q_{\rm ZA}+q_{\rm LA}=q_{\rm ZO}$. The energy conservation is analogous. 

Different from graphene, Fig.~\ref{fig_scattering}~(c) shows that there is no crossing between the TA/LA branches and the optical branches in MoS$_{2}$, because of the energy gap between acoustic and optical branches. This indicates that there is no permitted scattering for the ZA mode in MoS$_{2}$, which has the important implication that the resonant oscillation in the MoS$_{2}$ can occur for a long time with less intrinsic energy dissipation.  In other words, the energy gap between acoustic and optical branches in MoS$_{2}$ helps to prevent the resonant oscillation from being interrupted by other vibrational modes, and is why MoS$_{2}$ nanoresonators exhibit significantly less energy dissipation than graphene nanoresonators.

We note that a similar energy gap also exists in the phonon dispersion of other dichalcogenides like WS$_{2}$.\cite{DamnjanovicM2008mmp,SanchezAM} Based on the above discussion, these materials are also expected to have less intrinsic energy dissipation than graphene nanoresonators.

\subsection{Nonlinear and Strain Effects}

In addition to studying the differences in intrinsic energy dissipation, we now study additional effects that have previously been used to tailor, or enhance the resonant properties of graphene nanoresonators.  For example, recent studies have shown that inducing large, nonlinear oscillations of graphene may lead to an increased mass sensitivity.\cite{DaiMD,AtalayaJ2010epl,JiangJW2012nanotechnology}  Similarly, researchers have shown that the application of tensile mechanical strain can substantially increase the Q-factors, and thus the mass sensitivity of graphene.\cite{KimSYnanotechnology} The issue we consider now is the utility of these techniques on MoS$_{2}$ nanoresonators.

Fig.~\ref{fig_energy_alpha} shows that the energy dissipation can be affected by the actuation parameter $\beta$. Fig.~\ref{fig_energy_alpha} compares the kinetic energy time history in graphene and MoS$_{2}$ nanoresonators at T=4.2~K. It shows that the energy dissipation in both systems becomes stronger as the actuation parameter increases. This is because of the \jwj{nonlinear interaction between the oscillation mode and other vibration modes} in the graphene or MoS$_{2}$ induced by the large actuation parameter; i.e the graphene or MoS$_{2}$ nanoresonator is stretched so much in the sine-wave-like configuration that it can not contract back to its original length after it reaches the horizontal position. As a result, some obvious ripples occur in the configuration with maximum kinetic energy as shown in the two insets for $\beta=5$ panel. These two insets show two special configurations, which have minimum or maximum kinetic energy. A direct result from these ripples is the generation of other z-direction vibration modes, leading to the decoherence of the resonant oscillation. This decoherence effect results in stronger energy dissipation in the graphene and MoS$_{2}$ nanoresonators actuated by large $\beta$ parameter. \jwj{This is similar to the effect that initial slack has in degrading the Q-factors of graphene nanoresonators. Garcia-Sanchez et al. found that an initial slack leads to a specific vibrational mode that is localized at the free edges of the graphene nanoresonator. In these vibrations, graphene vibrates in the perpendicular direction; i.e in the same direction as the nanomechanical oscillation direction. As a result, the nanomechanical resonant oscillation of the graphene will be affected by these edge vibrations, leading to a lower Q-factor.\cite{SanchezDG}}

Fig.~\ref{fig_energy_strain} (left) shows that the energy dissipation in graphene can effectively be eliminated through the application of tensile mechanical strain.  These results were obtained for Fig.~\ref{fig_energy_strain} using the simulation parameters of T=50.0~K and $\beta=2.0$. For the graphene nanoresonator, the energy dissipation is minimized by applying tensile strains larger than $\epsilon=0.01$, or 1\%, which has previously been observed by Kim and Park.\cite{KimSY2009apl} The energy dissipation in MoS$_{2}$ nanoresonators can also be reduced by the tensile strain.(see right panel in Fig.~\ref{fig_energy_strain}), though we note that the Q-factors of graphene are more strongly enhanced by strain as they are significantly smaller without any applied strain.

To understand the effects of tensile strain on the energy dissipation, we monitor the structural evolution from the molecular dynamics simulation for both graphene and MoS$_{2}$. Fig.~\ref{fig_cfg_strain} shows two special configurations corresponding to minimum or maximum kinetic energy, for graphene (left) and MoS$_{2}$ nanoresonators (right). Panels (a)-(c) are configurations of graphene nanoresoantors with increasing mechanical tension. Panel (a) shows some obvious ripples (indicated by green ellipses) in the horizontal configuration, i.e with maximum kinetic energy. These ripples are the result of thermal vibrations at 50~K, and they occur because it is energetically much easier for graphene to bend than to deform in-plane. These ripples are smaller than those generated due to large actuation energy as shown in Fig.~\ref{fig_energy_alpha}.  As we have mentioned above, the function of ripples is to generate other z-direction vibrational modes, leading to stronger energy dissipation in the graphene nanoresonator. Panels (b) and (c) show that these thermal vibration induced ripples in the graphene nanoresonator can be completely eliminated by the mechanical tension, as the graphene nanoresonator recovers its original horizontal shape at its maximum kinetic energy state. That is the origin for the decreasing energy dissipation by tensile strain in graphene nanoresonators in Fig.~\ref{fig_energy_strain}. Fig.~\ref{fig_cfg_strain}~(d)-(f) are configurations for MoS$_{2}$ nanoresonators with increasing mechanical tension. The thermal vibration-induced rippling is effectively not observed in the MoS$_{2}$ nanoresonator, mainly due to its large bending modulus as compared with graphene.\cite{JiangJW2013bend} These small ripples are also completely eliminated by the mechanical strains we applied, leading again to reduced energy dissipation in MoS$_{2}$ as seen in Fig.~\ref{fig_energy_strain}.

\jwj{Finally, it should be noted that the system size in our simulation is substantially smaller than would typically be examined in experiments. An interesting open issue is to address the dimensional crossover in the MoS$_{2}$ nanoresonators, where the oscillation-induced local strain close to the clamped boundary is known to have an important effect.\cite{UnterreithmeierQP2010prl,BartonRA}}

\section{Conclusion}
In conclusion, we have utilized classical molecular dynamics simulations to compare the intrinsic energy dissipation in single-layer MoS$_{2}$ nanoresonators to that in single-layer graphene nanoresonators.  Our key finding is that the energy dissipation in MoS$_{2}$ nanoresonators is considerably less than in graphene nanoresonators for the same conditions, endowing MoS$_{2}$ with both higher Q-factors and figure of merit as compared to graphene nanoresonators. Based on the phonon-phonon scattering mechanism, we attribute the reduced energy dissipation in MoS$_{2}$ to the large energy gap in its phonon dispersion, which helps to prevent the resonant oscillation from being interrupted by other vibrational modes.  This energy gap in the phonon dispersion is also observed in other dichalcogenides, such as WS$_{2}$, which suggests that this class of materials may generally exhibit lower energy dissipation and higher Q-factors as nanoresonators.  We also demonstrate that nonlinear actuation leads to larger energy dissipation in MoS$_{2}$ as compared to graphene due to the existence of additional ripples in MoS$_{2}$, though tensile mechanical strain is effective in reducing the energy dissipation in both graphene and MoS$_{2}$.  

\textbf{Acknowledgements} The work is supported by the Recruitment Program of Global Youth Experts of China and the German Research Foundation (DFG). HSP acknowledge the support of the mechanical engineering department at Boston University.


%
\end{document}